\documentclass[referee]{aa}

\usepackage{graphicx}
\usepackage{natbib}
\usepackage{latexsym}
\usepackage{amsmath}
\usepackage{amssymb}
\usepackage{amstext}
\usepackage{color}
\usepackage{psfrag}

\def\k{km s$^{-1}$}
\def\ks{km s$^{-1}$~}
\def\d{$^\circ$}
\def\m{$^\prime$}
\def\s{$^{\prime\prime}$}
\def\hh{$^{\mathrm h}$}
\def\mm{$^{\mathrm m}$}
\def\ss{$^{\mathrm s}$}
\def\cm3{cm$^{-3}$}

\def\2{$^{12}$CO}
\def\3{$^{13}$CO}
\def\H{HCO$^{+}$}
\def\msol{M$_\odot$}

\begin{document}

\title{A molecular outflow evidencing star formation activity in the vicinity of the HII region G034.8-0.7 and
the SNR W44} 
\author {S. Paron \inst{1}
\and M. E. Ortega \inst{1}
\and M. Rubio  \inst{2}
\and G. Dubner  \inst{1}
}

\institute{Instituto de Astronom\'{\i}a y F\'{\i}sica del Espacio (IAFE),
             CC 67, Suc. 28, 1428 Buenos Aires, Argentina\\
             \email{sparon@iafe.uba.ar} 
\and Departamento de Astronom\'{\i}a, Universidad de Chile, Casilla 36-D, 
	Santiago, Chile}

\offprints{S. Paron}

   \date{Received <date>; Accepted <date>}

\abstract{}{This work aims at investigating the molecular gas component in the vicinity of two young 
stellar object (YSO) candidates identified at the border of the HII region G034.8-0.7 that 
is evolving within a molecular cloud shocked by the SNR W44.
The purpose is to explore signatures of star forming activity in this complex region.} 
{We performed a near and mid infrared study towards the border of the HII region G034.8-0.7 and 
observed a 90\s $\times$ 90\s~region near 18\hh 56\mm 48\ss, 
$+$01\d 18\m 45\s (J2000) using the Atacama Submillimeter Telescope Experiment (ASTE) 
in the \2 J=3--2, \3 J=3--2, \H~J=4--3 and CS J=7--6 lines with an angular resolution of 22\s.}
{Based on the infrared study we propose that the source 
2MASS 18564827+0118471 (IR1 in this work) is a YSO candidate. 
We discovered a bipolar \2 outflow in the direction of the line of sight and a \H~clump towards IR1, 
confirming that it is a YSO. 
From the detection of the CS J=7--6 line we infer the presence 
of high density ($>$10$^{7}$ cm$^{-3}$) and warm ($>$60 K) gas towards IR1, probably belonging to the protostellar 
envelope where the YSO is forming. We investigated the possible 
genetic connection of IR1 with the SNR and the HII region. By comparing the dynamical 
time of the outflows and the age of the SNR W44, we conclude that the possibility of the SNR has triggered the 
formation of IR1 is unlikely. On the other hand, we suggest that the expansion of the HII region G034.8-0.7 is 
responsible for the formation of IR1 through the ``collect and collapse'' process.} 
{}

\titlerunning{Star formation activity towards the SNR W44}
\authorrunning{S. Paron et al.}


\maketitle

\section{Introduction}

The formation of stars occurs deep inside molecular clouds. In the standard scenario for the formation 
of an isolated low-mass star, a cold core contracts as magnetic and turbulent support are lost and 
subsequently collapses to form a protostar with a surrounding disk. Then, a stellar wind breaks out 
along the rotational axis of the system and drives a bipolar outflow, which gradually disperses the 
protostellar envelope revealing a pre-main sequence star with a disk \citep{van08}. 

Several authors have studied the role of the supersonic turbulence in the star forming 
processes (e.g. \citealt{mac04}). They suggest that when stars form the process is quick and dynamic, 
with gravitational collapse occurring at a rate controlled by supersonic turbulence. The balance between 
turbulence and gravity provides a natural explanation for the widely varying star formation rates 
seen at both cloud and galactic scales. 

As the largest contribution to interstellar turbulence comes 
from supernova explosions, it is thought that the dominant driving mechanism in star-forming regions of 
galaxies appears to be supernovae, while elsewhere coupling of rotation to the gas through magnetic fields
or gravity may be important \citep{mac04}.

\begin{figure}[h]
\centering
\includegraphics[width=10cm]{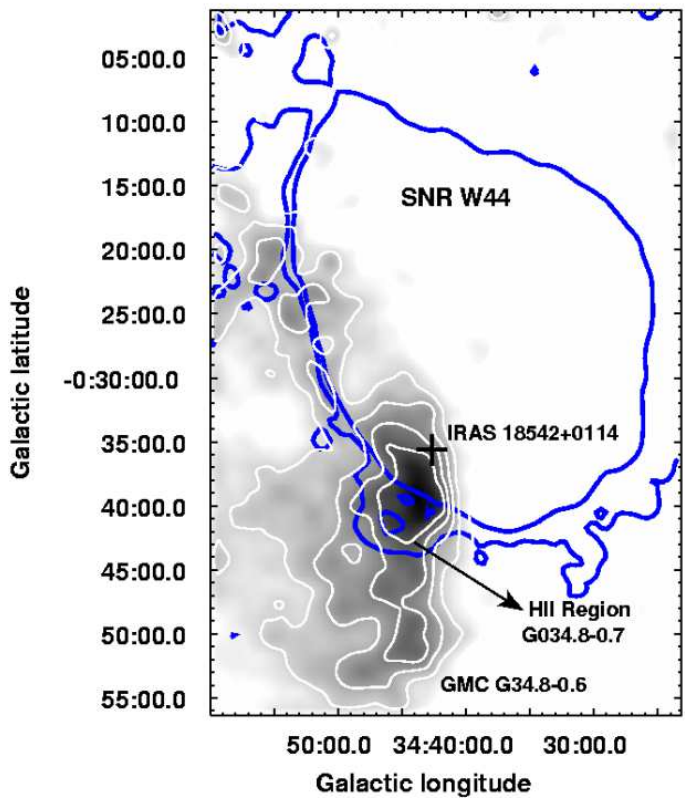}
\caption{\3 J=1--0 line averaged between 40 and 50 \ks showing the giant molecular cloud GMC G34.8-0.6. The thick blue 
contours outline the radio continuum emission of the SNR W44 and the HII region G034.8-0.7 at 20 cm. 
The cross shows the position of IRAS 18542+0114.}
\label{present}
\end{figure}

Since the early suggestion made by \citet{opik53} several numerical studies (e.g. \citealt{vanha98,melioli06})
proposed that supernova remnants (SNRs) can initiate the formation of new generations of stars. However, 
observational evidence of star formation triggered by SNRs is still scarce. 
The bright SNR W44 and the environment where it is evolving is a good target to explore this possibility. 
W44 is one of the few demonstrated cases of a SNR-molecular cloud interacting system. \citet{seta98} 
discovered six giant molecular clouds that appear to be partially surrounding the remnant. Later, \citet{seta04} 
proved that some of these clouds (at v$_{\rm LSR} \sim$ 45 \k)
are physically interacting with the remnant on its southeastern side. The CO spectra corresponding to these 
molecular structures present wing emission that, as 
the authors conclude, unambiguously confirm the interaction between W44 and the molecular gas.
The physical interaction of the blast wave with a clumpy interstellar medium is also supported by the detection 
of bright OH (1720 MHz) masers at LSR velocities approximately between 43 and 47 \k~(\citealt{hoffman05}, 
and references therein) and shocked H$_{2}$ emission observed with Spitzer-IRAC \citep{reach06}.
Many infrared sources are embedded in the portion of the giant molecular cloud that is being shocked by the SNR.
In particular IRAS 18542+0114, an IRAS point source
that coincides with the border of the HII region G034.8-0.7 \citep{kuchar97,ortega07} appears immersed in a region of 
shocked molecular gas interior to the SNR shell as seen in projection.  
Figure \ref{present} shows the molecular cloud 
in the \3 J=1--0 line (as extracted from the GRS\footnote{Galactic Ring Survey 
\citep{jackson06}}) from the average between 40 and 50 \k, the velocity 
range where the molecular cloud peaks, and also in coincidence with the systemic velocity of W44. The radio continuum of
the SNR and the HII region at 20 cm as extracted from the MAGPIS \citep{helfand06} are depicted in thick blue 
contours. 

The distance to W44 and the shocked 
molecular cloud was estimated to be $\sim$ 3 kpc \citep{rada72,caswell75,seta04}. Taking into 
account the recombination line velocity of 52.1$\pm$3.6 \ks derived by \citet{lockman89} and the HI absorption 
study performed by \citet{ortega07}, the same distance as the SNR is suggested for the HII region G034.8-0.7. 
In what follows we adopt 3 kpc as the distance for the SNR W44, the giant molecular cloud GMC G34.8-0.6 and 
the HII region G034.8-0.7.

IRAS 18542+0114 (cross in Figure \ref{present}) is resolved into several sources in the Two Micron All-Sky 
Point Source Catalog (2MASS, \citealt{cutri03}), and according to color criteria (see Section 3.1) some of them are 
young stellar object (YSO) candidates.

Thus, the scenario is that there are YSO candidates within a shocked molecular cloud located in a region
peripheric to a SNR and simultaneously on the border of an expanding HII region.

In this work we present a near and mid infrared study and molecular observations in the \2 and \3 J=3--2, \H~J=4--3 and 
CS J=7--6 lines towards IRAS 18542+0114 carried out with the purpose of exploring the star forming 
activity in the region and its connection with the local SNR and the HII region.

\section{Observations}

The molecular observations were performed on June 25, 2008 with the 10 m Atacama Submillimeter Telescope Experiment (ASTE; 
\citealt{ezawa04}). We used the CATS345 GHz band receiver, which is a two-single band SIS receiver remotely tunable 
in the LO frequency range of 324-372 GHz. We simultaneously observed \2 J=3--2 at 345.796 GHz and \H~J=4--3 at 
356.734 GHz, mapping a region of 90\s~$\times$ 90\s~centered at the position of IRAS 18542+0114 
(RA $=$ 18\hh 56\mm 47.8\ss, dec. $=$ $+$01\d 18\m 45\s, J2000). The mapping grid spacing was 10\s~and the integration
time was 72 sec. per pointing. Additionally we observed \3 J=3--2 at 330.588 GHz and CS J=7--6 at 342.883 GHz 
towards the centre of the region. All the observations were performed in position switching mode. The off position 
was RA $=$ 18\hh 55\mm 30\ss, dec. $=$ $+$01\d 39\m 55\s~(J2000) that was checked to be free of emission.

We used the XF digital spectrometer with a bandwidth and spectral resolution set to 128 MHz and 125 kHz, respectively.
The velocity resolution was 0.11 \ks and the half-power beamwidth (HPBW) was 22\s~at 345 GHz. The system temperature
varied from T$_{\rm sys} = 400$ to 700 K. The typical rms noise (in units of T$_{\rm mb}$) 
ranged between 0.1 and 0.4 K and the main beam efficiency was $\eta_{\rm mb} \sim 0.65$. 

The spectra were Hanning smoothed to improve the signal-to-noise ratio and only linear or/and some third order 
polinomia were used for baseline fitting. 
The spectra were processed using the XSpec software package developed at the Onsala Space Observatory.

\section{Results}

\subsection{Infrared emission}

\begin{figure}[h]
\centering
\includegraphics[width=11cm]{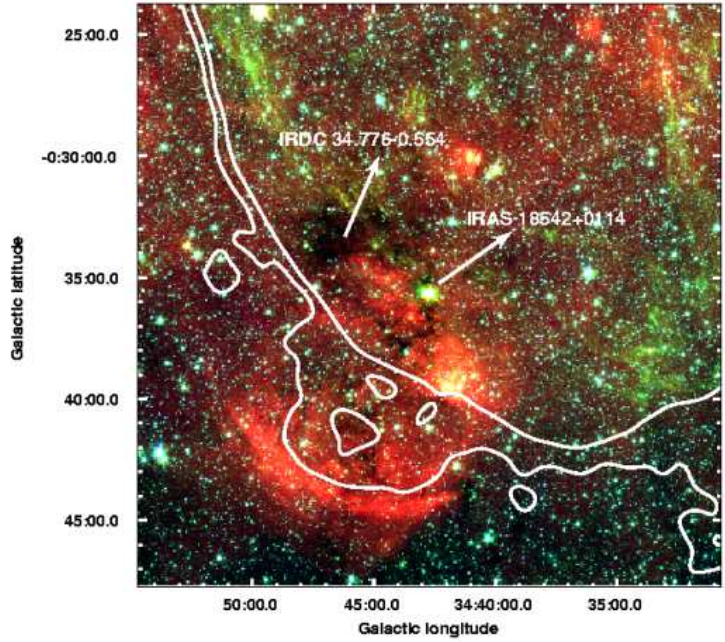}
\caption{{\it Spitzer}-IRAC three color image (3.5 $\mu$m $=$ blue, 4.5 $\mu$m $=$ green and 8 $\mu$m $=$ red)  
of a region about 22\m~$\times$ 22\m~around IRAS 18542+0114. The contours
correspond to the continuum emission at 20 cm of the SNR W44 and the HII region G034.8-0.7.}
\label{spitz}
\end{figure}

Figure \ref{spitz} shows a {\it Spitzer}-IRAC three color image of a region about 22\m~$\times$ 22\m~around 
IRAS 18542+0114 as extracted from GLIMPSE \citep{fazio04,werner04}. The three IR bands are 3.6 $\mu$m 
(in blue), 4.5 $\mu$m (in green) and 8 $\mu$m (in red). Contours of the radio continuum emission at 20 cm of 
the SNR W44 and the HII region G034.8-0.7 are included.

The IRAC 4.5 $\mu$m band contains both H$_{2}$ ($\nu =$ 0--0, S(9,10,11)) lines and CO ($\nu = 1-0$) band heads.
As noticed by \citet{cyga08}, all of these lines may be excited by shocks, such as those expected when protostellar 
outflows crash into the ambient ISM. In Figure \ref{spitz} it can be noticed that IRAS 18542+0114   
appears slightly extended in the 4.5 $\mu$m emission (green), therefore suggesting that it may be a YSO outflow candidate. 
It is important to note the difference between the emission from this YSO candidate
and from the gas shocked by the SNR, both seen in green. The first one is 
bright and concentrated around IRAS 18542+0114, while the second one is seen as diffuse filaments.
From Figure \ref{spitz} it is also evident the presence of an infrared dark cloud (IRDC 34.776-0554 
as catalogued in the Glimpse Dark Cloud Catalog). IRAS 18542+0114 is seen in projection over a border of 
this infrared dark cloud (IRDC). The IRDCs have been shown in recent years to be sites of the earliest stages 
of star formation \citep{rath06,rath07}.

\begin{figure}[h]
\centering
\includegraphics[width=8cm]{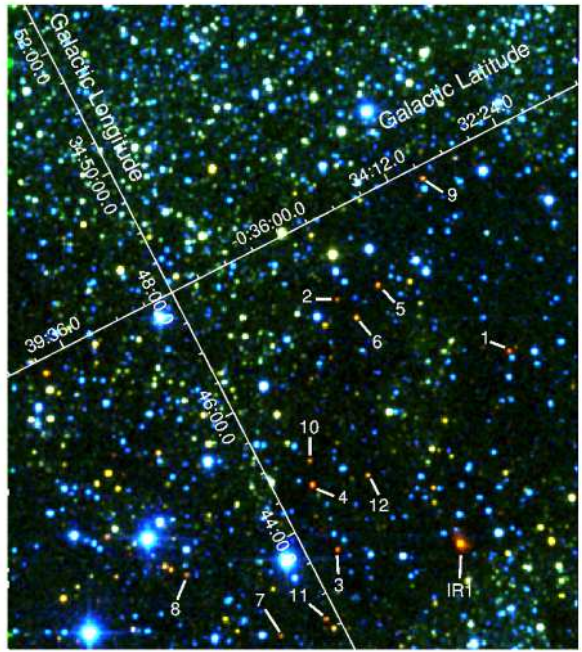}
\caption{2MASS {\it JHK} three-color image towards IRDC 34.776-0.554. The most reddened NIR sources 
are indicated with numbers. IRAS 18542+0114 is resolved into several 2MASS sources. It can be noticed that
the brightest and the most reddened one is 2MASS 18564827+0118471. }
\label{2mas}
\end{figure}

Figure \ref{2mas} shows near infrared (NIR) {\it JHK} three-color image in a region of about 10\m~$\times$ 10\m~that 
contains IRDC 34.776-0.554 obtained from the 2MASS Survey \citep{sk06}. The most reddened NIR 
sources are indicated with numbers. Among these sources it is included the bright 2MASS 18564827+0118471, 
possible counterpart of IRAS 18542+0114. 

\eject

\begin{figure}[h]
\centering
\includegraphics[totalheight=0.28\textheight,viewport=0 0 660 450,clip]{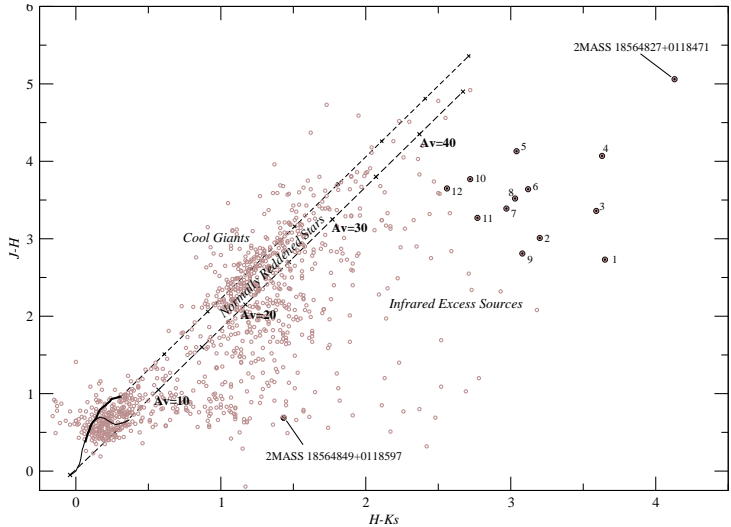}
\caption{Color-color diagram of the infrared sources in the vicinity of IRDC 34.776-0.554.  
The two solid curves represent the location of the main
sequence (thin line) and the giant stars (thicker line) derived from
\citet{bessell88}. The parallel dashed lines are reddening
vectors with the crosses placed at intervals corresponding to five magnitudes of visual extinction. 
We have assumed the interstellar reddening law of \citet{rieke85} 
($A_J/A_V$=0.282; $A_H/A_V$=0.175 and $A_K/A_V$=0.112). The plot is
classified into three regions: Cool Giants, Normally Reddened Stars and
Infrared Excess Sources. The most reddened sources are indicated as circled dots and 
the numbers correspond to the numbered sources of Figure \ref{2mas}. }
\label{cc}
\end{figure}

To look for primary tracers of stellar formation activity in the
vicinity of the IRDC, we used the 2MASS All-Sky Point Source Catalogue
in bands {\it J} (1.25 $\mu$m), {\it H} (1.65
$\mu$m) and {\it K} (2.17 $\mu$m). We performed photometry
in the region shown in Figure \ref{2mas}. 
Figure \ref{cc} shows the ({\it H-Ks}) versus ({\it J-H}) color-color (CC) diagram.
We selected sources with $Ks \leq$ 15 and with detection in at least two bands.  The typical 
color errors are about 5 \%.  
Based on the above criteria we found 1029 sources in the mentioned region.
Following \citet{hanson97}, we calculated for each source the parameter $q = (J-H)-1.83 \times (H-Ks)$, 
which determines the distance to the reddened vector.
YSO candidates, located at the Infrared Excess Sources region in Figure \ref{cc}, have 
$q$-values less than $-0.15$. From this criterium we found 364 YSO candidates.
The most reddened sources shown in Figure \ref{cc} as circled dots present 
the best quality detection in the H and K bands. The numbers
correspond to the numbered sources in Figure \ref{2mas}.
Their locations in the CC diagram indicate that they are likely YSOs,
where the NIR excess around protostars is due to the optically thick circumstellar disk/envelopes.
Figure \ref{cc} also shows two different groups of sources located between the parallel dashed lines. The 
first one, close to the origin of the diagram, corresponds to several blue sources which can be seen
in Figure \ref{2mas}. They are 
probably foreground stars not related to the IRDC, and as can be noticed in the CC diagram
they are clustered near the main sequence curve with visual extinction values 
from A$_{V} \sim 3$ to 8 mag. The second group is located between a visual extinction that 
ranges from A$_{V} \sim 18$ to 28 mag. Most of these sources correspond to background and 
possible molecular cloud embedded main sequence stars.

\begin{figure}[h]
\centering
\includegraphics[width=10cm]{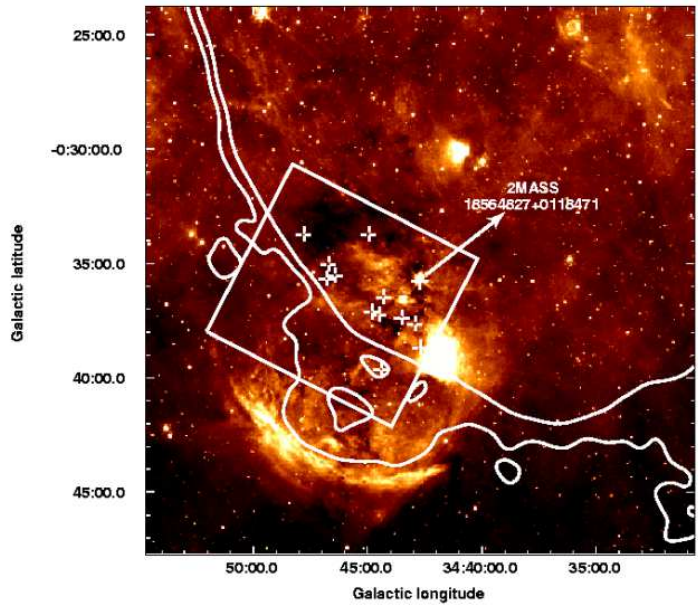}
\caption{IRAC 8 $\mu$m band emission obtained from GLIMPSE. The crosses indicate the most reddened sources 
obtained from the photometric study performed in the area indicated by the box. The box represents the 
same region as shown in Figure \ref{2mas} }.
\label{IR8mu}
\end{figure}

Figure \ref{IR8mu} displays the spatial distribution of the most reddened sources mentioned above over
the IRAC 8 $\mu$m band emission. The emission in this IR band clearly depicts the border of the HII region, 
with the IR radiation mainly originated in 
the policyclic aromatic hydrocarbons (PAHs). The figure also includes the contours of the radio continuum 
emission of the SNR and the HII region and a box that indicates the region where the photometric 
study was performed. From this figure, it is notorious that the most reddened sources lie on the border of the 
HII region that is seen
in projection interior to the SNR shell. Also the location of these sources delineate the border of
IRDC 34.776-0.554.

\begin{figure}[h]
\centering
\includegraphics[totalheight=0.22\textheight,viewport=0 0 660 480,clip]{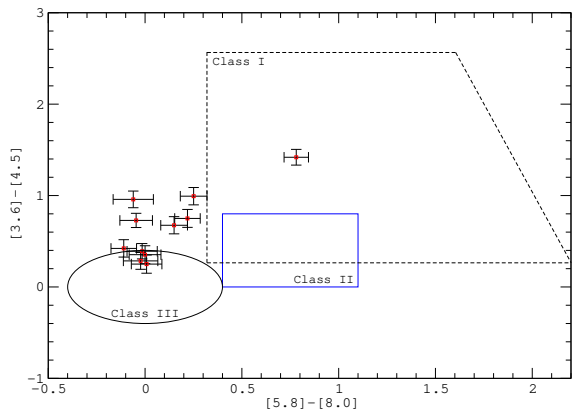}
\caption{GLIMPSE color-color diagram, [3.6] $-$ [4.5] versus [5.8] $-$ [8.0], for the most reddened sources remarked
in Figure \ref{2mas}. Class I and Class II regions are indicated following \citet{allen04}. The ellipse centered at 
0,0 encloses the region of main sequence and giant stars.}
\label{CCIRAC}
\end{figure}

Figure \ref{CCIRAC} shows IRAC color-color diagram for the most reddened sources highlighted and numbered 
in Figure \ref{2mas}. The regions that indicate the stellar evolutionary stages are based on the 
criteria presented by \citet{allen04}. 
The sources IR1 and 8 are not included because they have no detection in some IRAC bands. 
Only source 7 lies in the region of protostars with circumstellar envelopes (Class I), the sources 2, 5, 6, 10 
and 12 lie in the region of main sequence and giant stars (Class III). None source occupies the region of 
young stars with only disk emission (Class II). However the sources 1, 3, 4, 9 and 11, located outside the 
delimited regions, could therefore be reddened Class II objets \citep{allen04}.

In particular, in a region of 90\s~$\times$ 90\s~towards IRAS 18542+0114 (IR1), the region analyzed
in molecular lines (see next sections), the above 2MASS color criteria \citep{hanson97} identifies two 
sources: 2MASS 18564827+0118471, the most reddened source in the whole studied region (box in Figure 
\ref{IR8mu}), and 2MASS 18564849+0118597, both indicated in the CC diagram (Figure \ref{cc}). In what follows 
we analyze the molecular gas in this region.

\subsection{Molecular emission}

\subsubsection{\H~J=4--3}

\begin{figure}[h]
\centering
\includegraphics[width=8cm]{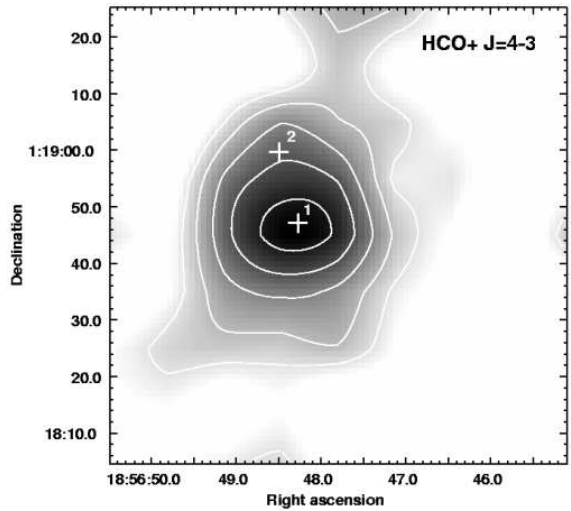}
\caption{\H~J=4--3 emission integrated between 40 and 50 \ks towards the infrared source IRAS 18542+0114. The 
contours levels are 3.5, 5, 7, 9 and 11 K \k. The crosses indicate the positions of the 2MASS point sources that
according to the color criteria described in the text could be YSOs. The angular resolution is $\sim$ 22\s~and
the $\sigma_{\rm rms} \sim 0.5$ K \k.}
\label{hco+I}
\end{figure}

Figure \ref{hco+I} shows a 90\s~$\times$ 90\s~map obtained towards the infrared source IRAS 18542+0114 in the 
\H~J=4--3 line integrated between 40 and 50 \k. As noticed above, in this region there are 
two 2MASS sources, YSO candidates: 
2MASS 18564827+0118471 (source 1 in Figure \ref{hco+I}, hereafter IR1) and 2MASS 18564849+0118597 
(source 2 in Figure \ref{hco+I}, hereafter IR2).

\begin{figure}[h]
\centering
\includegraphics[angle=-90,width=8cm]{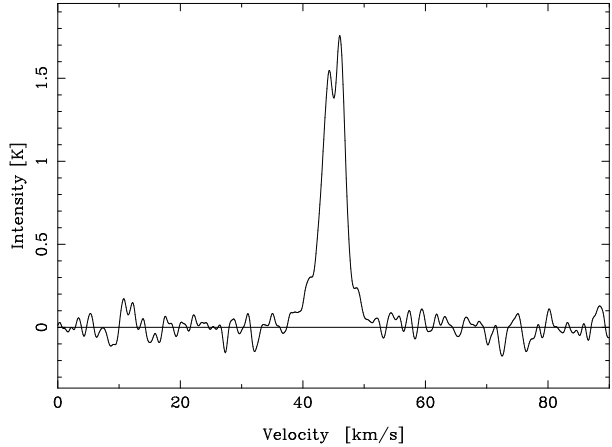}
\caption{\H~J=4--3 average profile of the region where emission is detected. The rms noise is $\sigma_{\rm rms} 
\sim 0.03$ K}
\label{hco+espec}
\end{figure}

From Figure \ref{hco+I} a dense \H~clump is evident with IR1 exactly lying over  
the \H~maximum and IR2 towards a border of the clump. By assuming as the limit of the 
\H~J=4--3 structure the 3.5 K \ks contour ($\sim 7$ times above the noise level), we can estimate a radius 
of $\sim$ 25\s~for this clump, 0.36 pc at the adopted distance of 3 kpc.

Figure \ref{hco+espec} shows an \H~J=4--3 spectrum obtained from the average over the whole observed feature.
From these observations we derived line parameters of the \H~J=4--3 that are presented in Table \ref{cloudparam}.
The average main-beam brightness
temperature $\langle$T$_{\rm mb}$$\rangle$, center line velocity, line width (FWHP), the average integrated
\H~intensity $\langle \int{T_{\rm mb}~dv} \rangle$ and the radius, are given in Cols. 1 to 5, respectively.
Errors are formal 1$\sigma$ value for the model of the Gaussian line shape.

\begin{table}[h]
\caption{Molecular clump parameters from the \H~J=4--3 emission.}
\label{cloudparam}
\centering
\begin{tabular}{ccccc}
\hline\hline
$\langle$T$_{\rm mb}$$\rangle$ & V$_{lsr}$ & $\Delta v$ & $\langle  \int{T_{\rm mb}~dv} \rangle$ & R  \\
(K) & (\k)   & (\k)     &   (K \k)  & (pc)              \\
\hline
1.65 $\pm$ 0.20 & 45.10 $\pm$ 0.30 & 4.40 $\pm$ 0.60 & 8.00 $\pm$ 1.00 & 0.36  \\
\hline
\end{tabular}
\end{table}

\begin{figure}[h]
\centering
\includegraphics[width=10cm]{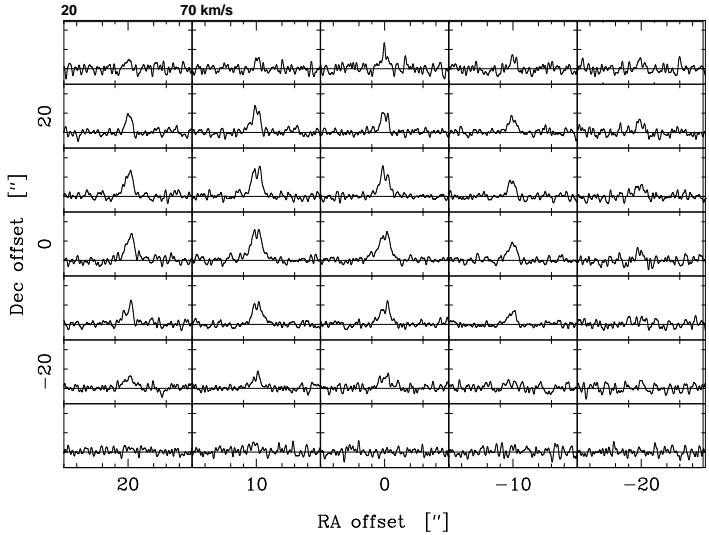}
\caption{\H~J=4--3 spectra obtained towards IR1, whose offset position is approximately (10\s, 0\s). The velocity range 
goes from 20 to 70 \ks and is shown at the top left profile. Note the profiles 
that present a dip at v $\sim 45$ \ks near the center of the surveyed region. }
\label{hco+profiles}
\end{figure}

The \H~J=4--3 profile presented in Figure \ref{hco+espec}  
presents a dip at v $\sim 45$ \k, the central velocity of the whole molecular complex. Only the profiles 
towards the center of the clump, that is towards IR1, clearly display this feature (see Figure \ref{hco+profiles}).
Such dip can be interpreted as a self-absorption effect caused by less excited gas. 
As \citet{hira07} propose, this kind of spectral features may indicate a significant population of the 
J = 3 level in the outer gas, that requires a moderately high density 
but with lower J=4--3 excitation temperature than in the center of the clump. Such effect might be illuminating 
the existence of a density gradient in the clump. 

We have also analyzed the existence of a possible blue asymmetry which would be an evidence 
of infalling gas towards a protostar \citep{leung77,zhou92}. In the star formation model of an infalling envelope, the 
blueshifted component is stronger than the redshifted one. This is due to the fact that the redshifted component 
is absorbed 
by the cooler infalling gas in the near half of the envelope along the line of sight. We did not find such an 
asymmetry in our spectra. However, as discussed by \citet{greg97,greg00}, the 
blue asymmetry line profiles do not appear in the totality of the known YSOs. One reason could be that if a redshifted 
outflow exists in the telescope beam, the redshifted components absorbed by the infalling envelope would be compensated.

\subsubsection{\2 J=3--2}

\begin{figure}[h]
\centering
\includegraphics[angle=-90,width=6cm]{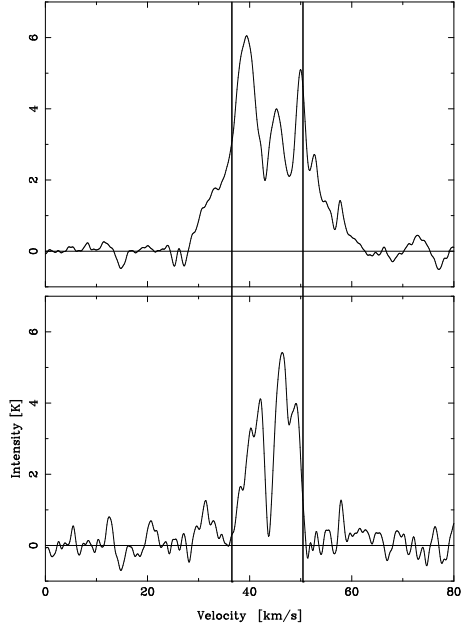}
\caption{\2 J=3--2 spectra. Upper: spectrum towards the center of the observed region. Bottom: spectrum 
towards the edge of the region. The borders between outflow and line core are shown with two vertical lines.
The rms noise is $\sigma_{\rm rms} \sim 0.15$ K}
\label{COespec}
\end{figure}

\begin{figure}[h]
\centering
\includegraphics[width=8cm]{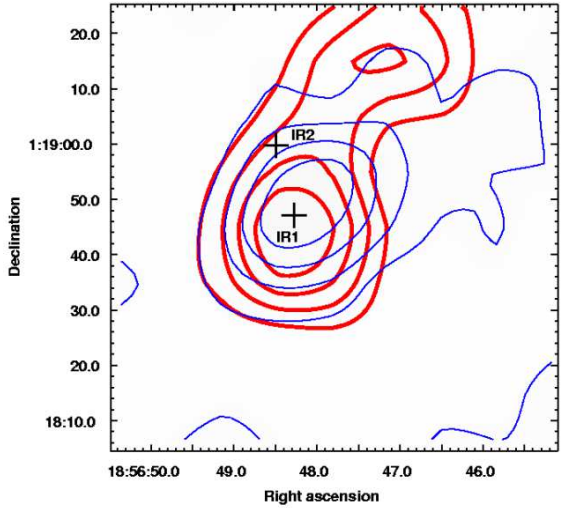}
\caption{The thin contours represents the \2 J=3--2 emission integrated between 27 and 37 \ks (blueshifted component). 
The contour levels are 5, 10, 15 and 20 K \k. The thick contours show the \2 J=3--2 emission integrated between 50 to 
63 \ks (redshifted component) and the levels are 10, 15, 20 and 25 K \k. The crosses indicate the positions of the 
2MASS sources IR1 and IR2.}
\label{out-cont}
\end{figure}

Analyzing the \2 J=3--2 data we find that the profiles towards IR1 are remarkably broader than 
those towards the border of the surveyed region. For comparison Figure \ref{COespec} shows two spectra: the upper one, 
obtained towards the center of the region and the bottom one towards the edge of the map. Such phenomena can be 
interpreted as evidence of an 
outflow observed along the line of sight driven by the source IR1. In effect, the blue component extends 
from 27 to 37 \ks while the red component goes from 50 to 63 \k. Figure \ref{out-cont} displays the \2 J=3--2 
emission integrated in both velocity ranges. 
The blue and red shifted components are presented in thin and thick contours, respectively. The crosses indicate 
the positions of the 2MASS sources IR1 and IR2.

It is worthwhile to remember that the \2 J=3--2 line almost always appears self-absorbed
towards star-forming regions and often reveals information on the gas kinematics \citep{johnstone03}.
Inspecting our data, most of the \2 profiles in the region have a dip at v $\sim 44$ \k.
This velocity is very close to the peak velocity of the \3 J=3--2 line (see subsection \ref{cs13}),
which is an optically thinner line. Such correspondence strongly suggests that the dip in the \2 profiles is indeed caused
by self-absorption by less excited gas (see e.g. \citealt{zhou93}). Thus we can discard the possibility of
two \2 emission components with different velocities along the line of sight.

\begin{figure}[h]
\centering
\includegraphics[width=14cm]{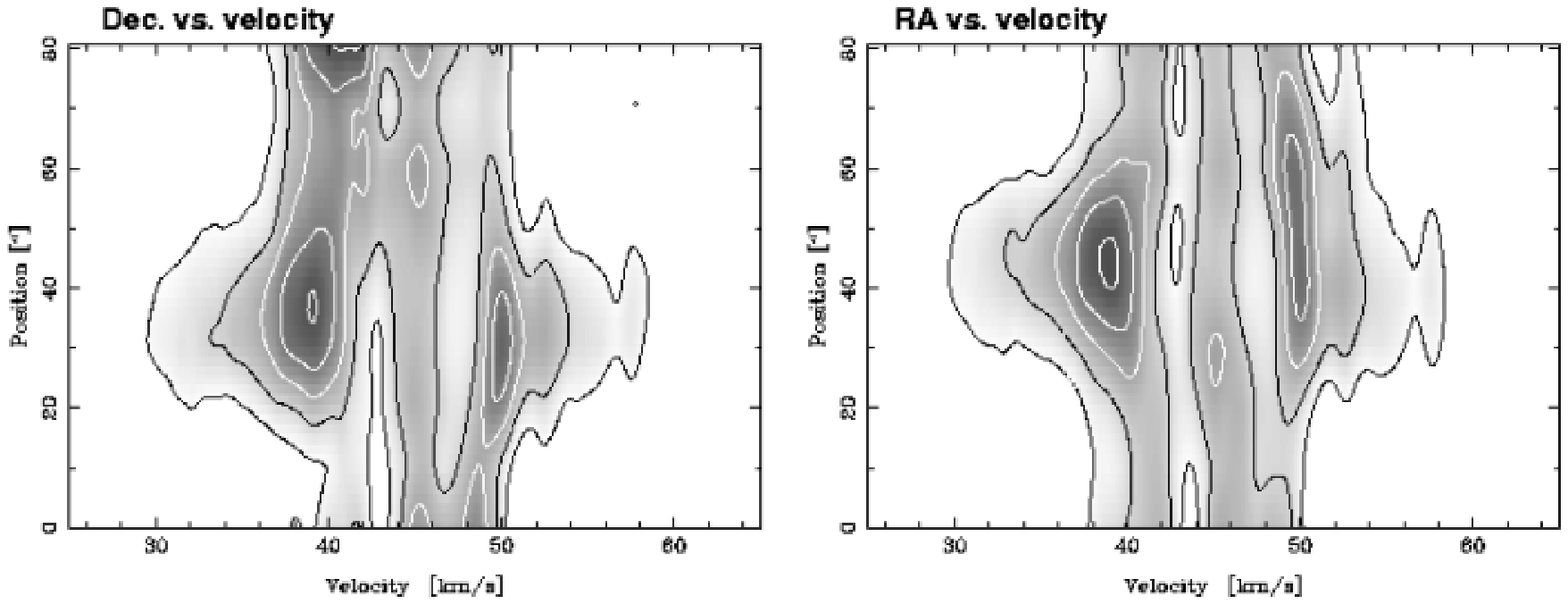}
\caption{\2 position-velocity diagrams. Left: along a line at constant RA $=$ 18\hh 56\mm 48\ss. 
Right: along a line at constant dec. $=$ $+$01\d 18\m 45\s.}
\label{velpos}
\end{figure}

Figure \ref{velpos} (left) displays the \2 position-velocity diagram along a line 
at constant Right Ascension (18\hh 56\mm 48\ss), while Figure \ref{velpos} (right) shows the same diagram along a line 
at constant Declination ($+$01\d 18\m 45\s). Redshifted and blueshifted features extending in velocity 
are clearly evident in both images.

\begin{figure}[h]
\centering
\includegraphics[width=7cm]{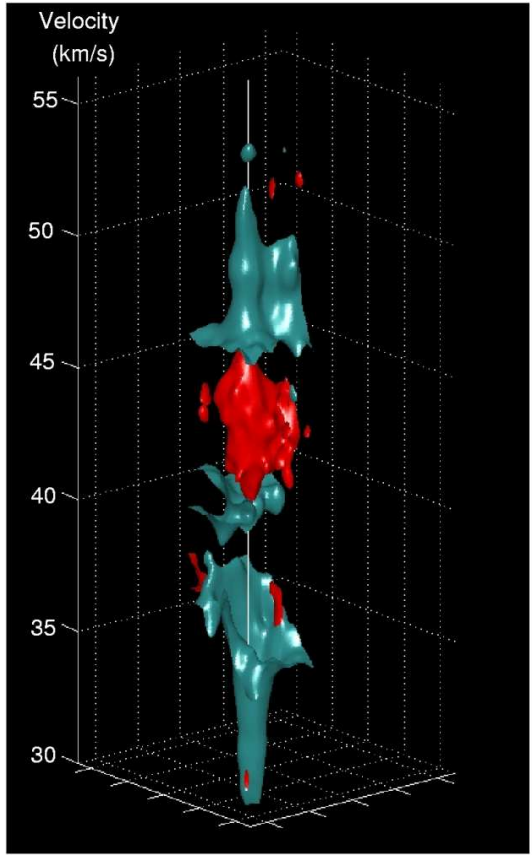}
\caption{Three dimensional picture showing isosurfaces of intensity. The 
feature between $\sim$ 40 and 45 \ks is the \H~emission at 0.7 K while the extended features in velocity 
(along the vertical axis) show the isosurface of the \2 emission at 1.8 K.}
\label{3d}
\end{figure}

Based on these observational results we can confirm that 
we are observing gas outflows originated in the source IR1.   
These results together with the above noticed slightly extended emission in the IRAC 4.5 $\mu$m 
band (indicative of shocked gas by outflows) and its position in the infrared color-color diagram (see Section 3.1), 
confirm that IR1 is a YSO.
In Figure \ref{3d} we present a three dimensional picture of the \H~structure and the outflows mapped by the \2.
The figure shows the \2 and the \H~data as isosurfaces
of intensity. The vertical axis corresponds to the velocity while the horizontal axes represent the spatial coordinates.
The feature at v $\sim$ 40 - 45 \ks is the \H~emission at 0.7 K. The extended features in velocity are the
\2 emission at 1.8 K, which clearly show the outflows structure.

\subsubsection{\3 J=3--2 and CS J=7--6} \label{cs13}

\begin{figure}[h]
\centering
\includegraphics[width=12cm]{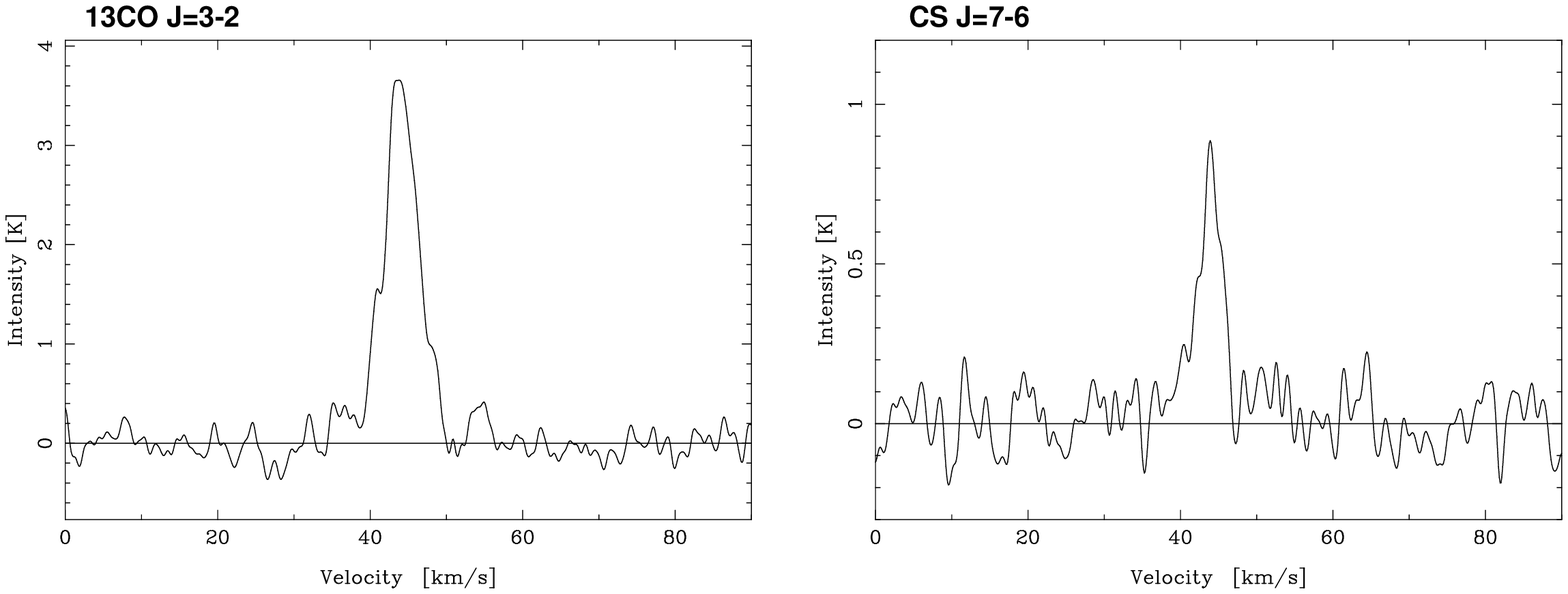}
\caption{\3 J=3--2 profile (left) and CS J=7--6 profile (right) obtained towards RA $=$ 18\hh 56\mm 47.8\ss, 
dec. $=$ $+$01\d 18\m 45\s~(J2000). The rms noise is $\sigma_{\rm rms} \sim 0.10$ and 0.07 K for each line, 
respectively.}
\label{13andCS}
\end{figure}

Figure \ref{13andCS} displays the profiles of the \3 J=3--2 and CS J=7--6 emission, left and right, respectively, 
obtained towards the center of the studied region (RA $=$ 18\hh 56\mm 47.8\ss, dec. $=$ $+$01\d 18\m 45\s, J2000).
The parameters determined from Gaussian fitting of these lines are presented in Table \ref{cs12tabla}.
T$_{mb}$ represents the peak brightness temperature, V$_{lsr}$ the central velocity,
$\Delta v$ the line width and $I$ the integrated line intensity. Errors are formal 1$\sigma$ value for the model 
of the Gaussian line shape.

\begin{table}[h]
\caption{Observed parameters of the \3 J=3--2 and CS J=7--6 emissions towards RA $=$ 18\hh 56\mm 47.8\ss,
dec. $=$ $+$01\d 18\m 45\s~(J2000).}
\label{cs12tabla}
\centering
\begin{tabular}{ccccc}
\hline\hline
Emission & T$_{mb}$ & V$_{lsr}$ & $\Delta v$  & $ I $    \\
         &  (K)     & (\k)      &   (\k)      & (K \k)                \\
\hline
\3 J=3--2 & 3.55 $\pm$0.30  &  44.10 $\pm$0.20 & 5.45 $\pm$0.50 & 21 $\pm$2 \\
\hline
CS J=7--6 & 0.50 $\pm$0.10  &  43.80 $\pm$0.90  & 4.00 $\pm$0.40 & 2.2 $\pm$0.5 \\
\hline
\end{tabular}
\end{table}

We have calculated the ratio of the integrated line intensities between the \3 J=3--2 and J=1--0 lines 
($^{13}$R$_{3-2/1-0}$). 
The latter was extracted from the GRS \citep{jackson06} and the J=3--2 line was convolved to the J=1--0 beam. 
We obtain $^{13}$R$_{3-2/1-0} \sim 1.4$. 

Besides, in order to compare the \3 J=3--2 with the \2 
J=3--2 to obtain an isotopic ratio between peak temperatures for the different velocity components, 
we have fit the \3 J=3--2 spectrum with three Gaussians.
We use the Gaussians with lower and higher velocities, closer to the blueshifted and redshifted components, respectively,
as seen in the \2 emission (see Figure \ref{COespec}). Thus we obtain the following peak temperature ratios:
$^{12}$T/$^{13}$T $\sim 13.5$ and $^{12}$T/$^{13}$T $\sim 8.8$ for the emissions close to the blueshifted and 
redshifted components, respectively. Finally, from these ratios we derive the optical depths for each line. 
For this region of the Galaxy, according to \citet{milam05}, we use an isotopic abundance 
ratio of [\2]/[\3] $\sim$ 70. Assuming an excitation temperature of T$_{ex}$ $\sim$ 20 K for both lines,
we derive the following optical depths: 
$\tau^{12} \sim 6$ and $\tau^{13} \sim 0.07$ for the emission close to the blueshifted component 
and $\tau^{12} \sim 10$ and $\tau^{13} \sim 0.1$ for the emission close to the redshifted one. 

\subsection{Kinematics and dynamics of the outflows}

In this section we investigate the physical parameters of the discovered outflows.
To carry out these calculations we applied the relations from \citet{choi93}. 
The CO column density ($N_{i}^{\rm CO}$) and the mass ($M_{i}$) for each channel are 
calculated from:

$$ N_{i}^{\rm CO} = 1.1 \times 10^{15} \frac{T_{3-2} \Delta v}{D(n,T_{k})} \frac{\tau_{32}}{1-exp(-\tau_{32})}~~{\rm and} $$

$$M_{i} = \mu m_{H_{2}} d^{2} \Omega N_{i}^{\rm CO} \frac{{\rm [H]}}{{\rm [C]}} \frac{{\rm [C]}}{{\rm [CO]}},$$
where $$ D(n,T_{k}) = f_{2}[J_{\nu}(T_{ex}) - J_{\nu}(T_{bk})][1 - exp(-16.597/T_{ex})], $$
$T_{3-2}$ is the peak temperature of the \2 J=3--2 line, $\Delta v$ the channel width, $f_{2}$ is the fraction of 
CO molecules in the J=2 state, $d = 3$ kpc is the distance to the giant molecular cloud,
$\Omega$ is the solid angle subtended by emission, [H]/[C] $= 2.5 \times 10^{3}$ \citep{greve91}, [C]/[CO] $= 8$
\citep{dickman78,van92}, and $\tau_{32}$ is 
the optical depth of the \2 J=3--2 transition calculated above, a value of 1.5 is adopted for $D(n,T_{k})$ following 
\citet{choi93}.
With these parameters we have estimated the mass $M$, the momentum $P$ and the kinetic energy $E_{k}$ for each outflow 
from: 
$$ M = \sum_{i}{M_{i}},$$
$$ P = \sum_{i}{M_{i} |v_{i} - v_{0}|},$$
$$ E_{k} = \sum_{i}{\frac{1}{2} M_{i} (v_{i} - v_{0})^{2}},$$
where $v_{0}$ is the systemic velocity, $\sim 45$ \k. Finally we calculate the dynamical timescale 
$t_{dyn} = R/V_{ch}$, where $V_{ch} = P/M$ is the characteristic velocity and $R$ is the size of the outflows, 
the mass loss rate $\dot{M} = M/t_{dyn}$ and the average driving force $F = P/t_{dyn}$. As the outflows are seen 
along the line of sight, that is the blue lobe is superposed on the red lobe, we estimate their sizes $R$ 
from the FWHM contour (see Figure \ref{out-cont}). The results are presented in Table \ref{outparam}.

\begin{table*}[h]
\caption{Outflows parameters.}
\label{outparam}
\centering
\begin{tabular}{lccccccc}
\hline\hline
Shift & $R$ & $M$ & $t_{dyn}$  & $\dot{M}$ & $P$ & $E_{k}$ & $F$    \\
         &  (pc)     & (\msol)      &   (yr)      & (\msol yr$^{-1}$) & (\msol \k) & (\msol [\k]$^{2}$) & (\msol \k yr$^{-1}$)\\
\hline
Blue & 0.35 & 7.0 $\times 10^{-4}$ & 2.6 $\times 10^{4}$ & 2.7 $\times 10^{-8}$ & 7.7 $\times 10^{-3}$ & 4.4 $\times 10^{-2}$ & 3.0 $\times 10^{-7}$ \\
\hline
Red  & 0.30 & 1.4 $\times 10^{-4}$ & 4.2 $\times 10^{4}$ & 3.3 $\times 10^{-8}$ &  1.2 $\times 10^{-2}$ & 5.5 $\times 10^{-2}$ & 2.7 $\times 10^{-7}$ \\
\hline
\end{tabular}
\end{table*}

\section{Discussion}

From the 2MASS data study we found that the sources with the largest NIR excess lie on the border 
of the HII region G034.8-0.7 and around of the dark cloud IRDC 34.776-0.554, a scenario that strongly suggests 
active star formation. The source 2MASS 18564827+0118471 (IR1 in this paper) is the most reddened source in the NIR 
and is also bright and slightly extended in the IRAC 4.5 $\mu$m band. This latter emission, 
which contains H$_{2}$ and CO vibrational lines, can be produced when protostellar outflows crash 
into the ambient ISM, suggesting that IR1 could be a YSO.

Our molecular observations have revealed the existence of a
\H~clump where IR1 is embedded. It is known that such molecular specie enhances 
in molecular outflows \citep{raw04}. In effect, a strong enhancement of the \H~abundance is 
expected to occur in the boundary layer between the outflow jet and the surrounding
molecular core. This would be due to the liberation and photoprocessing by the shock of the molecular material 
stored in the icy mantles of the dust. 
Indeed, through the \2 J=3--2 line we discovered the presence of outflows extended along the line of sight related to 
the \H~clump and IR1. In this context, we conclude that the source IR1 is a YSO driving bipolar outflows which are 
enhancing the abundance of \H.

The obtained outflows masses are similar to those obtained
for the outflows of the source IRS6, a low-mass YSO in Cederblad 110 \citep{hira07}.
We compared the derived parameters for the outflows (see Table \ref{outparam}) with those of 
the statistical study carried out by \citet{wu04} 
based on the properties of a sample of about 391 molecular outflows. In that work the sources 
were divided into outflows with low mass and outflows with high mass according to either the available bolometric 
luminosity of the central source or the outflow mass. Based on the compiled data the authors conclude
that the outflow phenomenon is common either in low and high mass star formation regions.
The red and blue outflows masses that we derived for IR1 are 
slightly smaller than the lowest masses discussed by \citet{wu04}. Therefore, 
IR1 would correspond to the ``low mass group'' as catalogued by them. It is important to remark that since 
we made no corrections to the velocities for possible projection effects, the momenta and kinetic energies that
we derive are lower limits. Also, as \citet{choi93} note, these effects can increase (if the flow is in the plane of 
the sky) or decrease (if the flow is along the line of sight) the timescale ($t_{dyn}$ in Table \ref{outparam}) by 
substantial factors. In addition, this time would be smaller if we use the maximum velocity rather than $V_{ch}$.
However our $t_{dyn}$ values are very close to the average dynamical time of the ``low mass group'' presented in 
\citet{wu04}. Other physical parameters, like kinetic energy, momentum and driving force listed in 
Table \ref{outparam} also agree with those of the ``low mass group''. We can therefore conclude that we 
are probably observing a source with outflows of low mass.

One stage in the formation of a star is the main accretion phase. This is the phase during which
the central object builds up its mass from a surrounding infalling envelope and accretion disk. Observational
evidence shows that the main accretion phase is always accompanied by a powerful ejection of a small fraction of
the accreted material in the form of prominent bipolar outflows (e.g. \citealt{mckee07,bachiller96}). Class $0$
protostars are considered the earlier stage of star formation with typical ages of
$1-3 \times 10^{4}$ years. Most of them, if not all, drive powerful ``jet-like'' CO molecular outflows, which
gradually disappear as the protostar goes into the following evolutionary stages. Thus, the dynamical times of
the outflows can be used as an approximation of the life time of YSOs. Therefore we suggest an age of a few
$\times 10^{4}$ years for IR1.

Additionally, we performed a Spectral Energy Distribution (SED) fitting of IR1 near and mid IR fluxes extracted from
the 2MASS Point Source Catalog (J, H and K bands), from the Glimpse Catalog ({\it Spitzer}-IRAC bands) and 
from the MSX Point Source Catalog (8.28, 12.13, 14.65 and 21.30 $\mu$m bands). 
Based on the SED models\footnote{http://caravan.astro.wisc.edu/protostars} 
presented by \citet{robi06,robi07} and considering a visual extinction between 15 and 30 mag (see Section 3.1), 
we suggest that IR1 can be a high mass YSO of $\sim 20$ \msol. Besides, 
the models predict an age between $7 \times 10^{4}$ and 10$^{5}$ years for this YSO, and a 
circumstellar disk of 0.01 \msol~with an accretion rate of $\sim 10^{-6}$ \msol~yr$^{-1}$.
The outflows and accretion rates presented in Table \ref{outparam} and derived from the SED models, respectively, 
are considerably lower than the typical rates of massive YSOs ($10^{-4}$ \msol~yr$^{-1}$ in both cases; \citealt{mckee03}). 
As noticed above, IR1 presents a high NIR excess, which suggests that it is embedded 
in a very dense envelope. Taking into account the YSO mass obtained from the SED models, the estimated ages
and the mentioned outflow and accretion rates for IR1, we suggest that this source is probably a massive YSO that 
could be near the end of the accretion process and has not ionized and dissipated its envelope yet.

The integrated line \3 3--2/1--0 ratio ($^{13}$R$_{3-2/1-0}$) obtained towards the center of the region is very 
similar to that obtained towards a circumstellar disk around a pre-main sequence star by \citet{vanZ}. As the 
authors remark, the $^{13}$R$_{3-2/1-0}$ could indicate temperatures of $\sim 20-40$ K. 
However, care should be taken with the interpretation of this result since the emission of the two lines most 
likely comes from different regions of the disk due to the difference in optical depth of the
two lines. 

The sulphur-bearing species, like the CS, have been associated with shocks and/or outflows. However it is still 
debatable whether sulphur-bearing species are good diagnostics of outflows \citep{johnstone03}. 
According to \citet{mori95}, the CS J=7--6 line traces high densities ($>$10$^{7}$ cm$^{-3}$) and warm 
temperatures ($>$60 K) in the protostellar envelopes.
Thus, we propose that the detection of this transition indicates the presence of dense and warm gas, 
showing the inner denser and warmer part of the protostellar envelope where IR1 is forming. 

\subsection{Could the SNR W44 or the HII region G034.8-0.7 have triggered the formation of IR1?}

From our molecular results and the IR emission study, we conclude that IR1 is indeed a YSO. 
As it lies in the border of the HII region G034.8-0.7, which is in turn evolving in a molecular cloud shocked by the 
SNR W44, we investigate their genetic connection.

According to the derived outflows dynamical time (see Table \ref{outparam}) and to the SED age estimate, 
IR1 might be a YSO with an age 
of a few $\times 10^{4}$ or maybe $\sim 10^{5}$ years. Based on the age of the associated 
pulsar, \citet{wol91} estimated an age of about $2 \times 10^{4}$ years for the SNR W44. Taking into account 
that the ages of IR1 and W44 are comparable and that a delay between the main triggering agent and the subsequent 
star formation is expected, it is unlikely that the SNR has triggered the formation of IR1.

On the other hand, as mentioned above, IR1 is seen in the plane of the sky overlapping  
the photodissociation region (PDR) associated with the HII region G034.8-0.7, located 
at the same distance that W44 \citep{ortega07}. 
In the literature there are many works about star formation triggered by the expansion of HII regions 
(e.g. \citealt{elme77,deha03,deha05,comeron05,zav06}). In what follows we investigate this possibility.  
From the \2 J=1--0 study performed by \citet{seta04}, a dense molecular shell surrounding the HII region G034.8-0.7
is evident at the velocity range 35 - 40 \k, which suggests that in this region can take place the ``collect and 
collapse'' process \citep{deha05,zav06}. In this process, the compressed shocked layer generated in the inhomogeneous 
medium by the expansion of the HII region may become gravitationally unstable along its surface on a long timescale. 
The process produces massive fragments of material, which allow the formation of massive stars and/or clusters.
According to \citet{ortega07}, taking into account that the exciting star of the HII region G034.8-0.7 should be of 
spectral type between O9.5 and O3, the age of the HII region would be 10$^{5}$ -- $3 \times 10^{6}$ 
years. These times are compatible with star forming processes on the associated PDR. 
Besides, as noticed in Section 3.1, the most reddened sources in the analyzed region lie 
preferentially in the border of the HII region that is seen in projection interior to the W44 shell.
We can therefore conclude that an scenario where the expansion of the HII region G034.8-0.7 has triggered 
the formation of IR1 is more possible.
Careful modelization of this scenario, including the actions of the HII region and the SNR, 
is planned for the near future.

Future molecular observations are planned to be performed towards this region in order 
to look for more YSOs embedded in the perturbed molecular cloud.
Besides, extended CS observations towards IR1 would be important. The transitions of this molecular specie are very 
useful to derive detailed density structure within the molecular envelopes of star-forming region.

\section{Summary}

We present an infrared study and molecular observations towards the IR source 
IRAS 18542+0114 carried out with the purpose of exploring signatures of star forming activity in  
the border of the HII region G034.8-0.7, which evolves within a molecular cloud shocked by the SNR W44.
We found that IRAS 18542+0114 presents a slight extended emission in the IRAC 4.5 $\mu$m band, which suggests 
that this source may be a YSO driven outflows. IRAS 18542+0114 is resolved into several sources 
in the 2MASS All-Sky Point Source Catalog, and two of them, according to color criteria are YSO candidates. 
One of these candidates, 2MASS 18564827+0118471 (IR1 in this work), is the most reddened source in the NIR 
in the studied region. From SED models we suggest that IR1 is probably a massive YSO that could be near the end of
the accretion process. Our molecular results confirms that IR1 is indeed a YSO. These results can be 
summarized as follows:

(a) We discovered a \H~clump towards IR1. It is known that such molecular specie enhances
in molecular outflows, which are strong evidences of star forming activity.  

(b) We discovered \2 outflows towards IR1 extended along the line of sight. 
From the dynamical time of these outflows and a SED age estimate, we suggest an age of a few $\times 10^{4}$ 
or maybe $\sim 10^{5}$ years for IR1. The discovery of molecular outflows confirms that IR1 is a YSO.

(c) From the detection of the CS J=7--6 line we confirm the presence of high density 
($>$10$^{7}$ cm$^{-3}$) and warm ($>$60 K) gas towards IR1, probably belonging to the protostellar envelope where 
this YSO is forming.

(d) By comparing the estimated age for IR1 and the age of the SNR W44 ($2 \times 10^{4}$ years), 
we conclude that it is unlikely that the SNR has triggered the formation of IR1. On the 
other hand, taking into account the age of the HII region G034.8-0.7 and that IR1, as other reddened sources, lie on 
its border, we find that the HII region may have initiated the star formation through the ``collect and collapse'' 
process. 

Future observations and theoretical modelization are planned in order to further investigate the star formation 
in this complex region.

\begin{acknowledgements}
We wish to thank the referee Dr. Zavagno whose constructive criticism has helped to improve the paper.
S.P. is grateful to the staff of ASTE for the support
received during the observations, especially to Juan Cort{\'e}s. Also S.P. acknowledges the 
support of Viviana Guzm{\'a}n during the observations.
S.P. is a postdoctoral fellow of CONICET, Argentina. M.O. is a doctoral fellow of CONICET, Argentina. 
G.D. is member of the {\sl Carrera del 
Investigador Cient\'\i fico} of CONICET, Argentina. This work was partially supported by the CONICET 
grant 6433/05, UBACYT A023 and ANPCYT PICT 04-14018.
M.R. is supported by the Chilean {\sl Center for Astrophysics}
FONDAP No. 15010003. M.R. and S.P. aknowledge support from FONDECYT N\d~1080335. 

\end{acknowledgements}

\bibliographystyle{aa}  
\bibliography{bib-W44}
\IfFileExists{\jobname.bbl}{}
{\typeout{}
\typeout{****************************************************}
\typeout{****************************************************}
\typeout{** Please run "bibtex \jobname" to optain}
\typeout{** the bibliography and then re-run LaTeX}
\typeout{** twice to fix the references!}
\typeout{****************************************************}
\typeout{****************************************************}
\typeout{}
}

\end{document}